\documentclass[submission,copyright,creativecommons]{eptcs}
\providecommand{\event}{SOS 2007} % Name of the event you are submitting to
\usepackage{breakurl}             % Not needed if you use pdflatex only.

\usepackage{graphicx}
\usepackage{amssymb}
\newtheorem{theorem}{Theorem}[section]

\newcommand{\Comment}[1] {}
\newcommand{\triple}[3]{\mbox{\ensuremath{\langle #1 \rangle~#2~\langle #3 \rangle}}}
\newcommand{\project}[2]{\ensuremath{{#1 \uparrow #2}}}
\newcommand{\transits}[3]{\ensuremath{#1\stackrel{#2}{\longrightarrow}#3}}
\newcommand{\lang}[1]{\ensuremath{{{\mathcal{L}}\left( #1 \right)}}}
\newcommand{\ie}{{\it i.e.\/}}

\newcommand{\dnote}[1]{\textit{Dimitra: #1}}

\newcommand{\Preds}{\mathsf{Preds}}
\newcommand{\abst}{\alpha}
\title{Abstraction and Learning for Infinite-State\\ Compositional Verification }
\author{Dimitra Giannakopoulou
\institute{NASA Ames\\
Moffett Field, CA, USA}
\email{dimitra.giannakopoulou@nasa.gov}
\and
Corina S. P\u{a}s\u{a}reanu
\institute{CMU/NASA Ames\\
Moffett Field, CA, USA}
\email{corina.s.pasareanu@nasa.gov}
}
\def\titlerunning{Infinite-State Compositional Verification}
\def\authorrunning{D. Giannakopoulou and C. S. P\u{a}s\u{a}reanu}
\begin{document}
\maketitle

\begin{abstract}
Despite many advances that enable the application of model checking techniques to the verification of large systems, the state-explosion problem remains the main challenge for scalability. Compositional verification addresses this challenge by decomposing the verification of a large system into the verification of its components. Recent techniques use learning-based approaches to automate compositional verification based on the assume-guarantee style reasoning. However, these techniques are only applicable to finite-state systems. In this work, we propose a new framework that interleaves abstraction and learning to perform automated compositional verification of infinite-state systems. We also discuss the role of learning and abstraction in the related context of interface generation for infinite-state components.
\end{abstract}

\section{Introduction}
\label{sec:introduction}

Despite several breakthroughs that enable the
application of model
checking to the verification of realistic systems, the essential
challenge in model checking remains the well-known state-space
explosion problem:
the size of a concurrent
reactive system to be verified is the product of the sizes
of its constituent components. Thus the cost of exhaustive verification as
done in model checking grows exponentially in the number of
state variables.

Compositional techniques attempt to tame this problem by applying
verification to individual components and merging the results without
analyzing the whole system.
%so that the combination avoids exponential explosion.
In checking components individually, it is often necessary to
incorporate some knowledge of
the context in which each component is expected to operate correctly.
Assume-guarantee reasoning~\cite{jones83a,pnueli84} addresses this
issue by using {\em assumptions} that capture the expectations that a
component makes about its environment.
The simplest such rule checks if a system composed of components $M_1$
and $M_2$ satisfies a property $P$ by checking that $M_1$
under assumption $A$ satisfies $P$ ({\em Premise~1}) and
discharging $A$ on the environment $M_2$ ({\em Premise~2}). For safety properties, %that are the focus of our paper,
{\em Premise~2} amounts to checking that $A$ is a conservative abstraction of $M_2$, \ie, an abstraction that preserves all of $M_2$'s execution paths.
This rule is also represented as follows, where the notation is described in
 more detail in Section~\ref{sec:formalisms}.

\[
\begin{array}{ll}
1: & \triple{A}{M_1}{P}\\
2: & \triple{\emph{true}}{M_2}{A}\\
\hline
   & \triple{\emph{true}}{M_1 \parallel M_2}{P}
\end{array}
\]

Assumptions have traditionally been developed manually, a limiting
factor to the practical impact of assume-guarantee reasoning.  In this
article, we review learning-based frameworks that automate the application
of assume-guarantee reasoning~\cite{cobleigh03,DBLP:journals/fmsd/PasareanuGBCB08}. 
These frameworks use the L* automata learning algorithm~\cite{angluin87} 
to generate and refine assumptions,
based on results obtained from model checking individual 
components separately.
Other related approaches followed (see e.g. articles in journal issue~\cite{DBLP:journals/fmsd/GiannakopoulouP08}).
\Comment{concentrated on ensuring minimality of the generated 
assumption for automatic application of the rule {\sc ASym}.
These learning-based techniques reduce this problem to that the 
computation of minimal separating automata~\cite{DBLP:conf/tacas/ChenFCTW09,DBLP:conf/cav/GuptaMF07} . The complexity of the resulting approach 
is higher that the original one that does not guarantee such minimality.
}

%This work uses the
%  observation~\cite{DBLP:conf/cav/GuptaMF07} that an assumption $A$
%  that satisfies both premises of the rule on the one hand includes
%  all traces of $M_1$, and on the other hand, is disjoint from the
%  traces of $M_2$ that violate $P$. In this sense, $A$ separates the
%  languages $L(M_1)$ and $L(M_2) \cap \overline{L(P)}$, where $L(M_i)$
%  describes the set of all traces of $M_i$, and similarly for $L(P)$.
 
%The second approach automates assume-guarantee reasoning by
%iteratively computing assumptions as abstractions of system
%components~\cite{agar}.  
These prior approaches were performed in the context of {\em finite-state} components.
In contrast, the focus of the present work is on compositional verification in
the context of {\em infinite-state} (or very large) components, where
the verification of individual components itself is intractable. We
introduce component abstractions for this task, and explain how
component abstraction refinement interacts with assumption generation.
%We explain the common principles between the learning-based and
%abstraction-based assume-guarantee reasoning for infinite-state
%systems, and then describe the details of each framework.
This article contributes a framework for
automated assume-guarantee reasoning of infinite-state systems.
%and explains their theoretical foundations. 
A related problem that we
studied in the past in the context of infinite-state systems is 
component interface generation~\cite{cav2010}. We explain here 
the differences between the two approaches, in particular the different types
of abstractions that are needed in each.
  
%The rest of the paper is organized as follows...

\section{Formalisms}
\label{sec:formalisms}

\subsection{Component Models and Properties}

%Here we need to decide what to use for models. 
%Show also safety properties and discuss parallel composition.

\subsubsection{Communicating State Machines}
\label{sec:fsms}

We model software components as (possibly infinite) communicating
state machines. Note that typically components have implicit finite
representations (e.g. a program) and only their semantics are given as
infinite state machines. %However, this is not important here.

Let $\mathcal{A}ct$ be the universal set of observable actions and let
$\tau$ denote a local action {\em unobservable} to a component's
environment. 
%Let $\pi$ denote a special \emph{error state}, which
%models safety violations; the error state has no outgoing transitions.

A CSM $M$ is a four-tuple $\langle Q,\alpha M, \delta, q_0\rangle$ where:
\begin{itemize}
\item $Q$ is a non-empty set of states.
\item $\alpha M\subseteq \mathcal{A}ct$ is a set of observable actions called the {\em alphabet} of $M$.
\item $\delta  \subseteq Q \times (\alpha M \cup\{\tau\}) \times Q$ is a
  transition relation
\item $q_0\in Q$ is the initial state
%\item $F \subseteq Q$ is a set of accepting states
\end{itemize}

%{\bf do we need an extra condition here?}
We write $q\stackrel{a}{\longrightarrow} q'$ for $(q,a,q')\in\delta$.
A \emph{trace} $t$ of a CSM $M$ is a finite sequence of observable
actions that label the transitions that $M$ can perform starting at
its initial state (ignoring the $\tau$-transitions).
%For an CSM
%$M$ and a trace $t$, let $\hat\delta(q, t)$ denote the set of states
%that $M$ can reach after reading $t$ starting at state $q$. A trace
%$t$ is said to be {\em accepted} by an CSM $M = \langle Q,\alpha M,
%\delta, q_0, F \rangle$ if $\hat\delta(q_0, t) \cap F \neq
%\emptyset$. 
The {\em language} of $M$, denoted $\lang{M}$ is
the set of all traces of $M$.
% $\{ t \mid \hat\delta(q_0, t) \cap F \neq \emptyset \}$.

Note that we assume that the state set
$Q$ may be infinite but the alphabet $\alpha M$
is finite; the actions in the alphabet can be seen as modeling 
e.g. sending and receiving of messages or method calls and returns
for software components. Method parameters are not treated here 
(see~\cite{DBLP:conf/cav/BeyerHS07,DBLP:conf/sas/GiannakopoulouRR12} for approaches that
handle parameters).

We sometimes abuse the notation and denote by $t$ both a trace and its
trace CSM.  For a trace $t$ of length $n$, its trace CSM consists of
$n+1$ states, where there is a transition between
states $m$ and $m+1$ on the $m^{th}$ action in the trace $t$.
For $\Sigma \subseteq \mathcal{A}ct$, we use $\project{t}{\Sigma}$ to
denote the trace obtained by removing from $t$ all occurrences of
actions $a \notin \Sigma$. Furthermore, $\project{M}{\Sigma}$ is defined
to be a CSM over alphabet $\Sigma$ which is obtained from $M$ by
renaming to $\tau$ all the transitions labeled with actions that are
not in $\Sigma$.  Let $t$, $t'$ be two traces and $\Sigma$, $\Sigma'$
be the sets of actions occurring in $t$, $t'$, respectively. By the
{\em symmetric difference} of $t$ and $t'$ we mean the symmetric
difference of the sets $\Sigma$ and $\Sigma'$.

A CSM $M = \langle Q,\alpha M, \delta, q_0\rangle$ 
is {\em non-deterministic} if any of these two conditions holds: 
1) it contains
$\tau$-transitions or 2) if there exists $(q, a, q'), (q, a, q'') \in \delta$
such that $q' \neq q''$. Otherwise, $M$ is \emph{deterministic}.

\subsubsection{Parallel Composition of CSMs} 

Let $M_1 = \langle Q^1, \alpha M_1, \delta^1, q_0^1\rangle$ and
$M_2 = \langle Q^2, \alpha M_2, \delta^2, q_0^2\rangle$ be two
CSMs. The parallel composition operator $\parallel$ is a commutative
and associative operator that combines the behavior of two components
by synchronizing the actions common to their alphabets and
interleaving the remaining actions. Formally, $M_1 \parallel M_2$ is
a CSM $M = \langle Q,\alpha M, \delta, q_0\rangle$, 
where $Q = Q^1 \times Q^2$, $q_0 = (q^1_0,
q^2_0)$, $\alpha M = \alpha M_1 \cup \alpha M_2$, and $\delta$ is
defined as follows, where $q_1, q_1'\in Q^1$ and $q_2, q_2'\in Q^2$: 
% (note that the symmetric rules are implied by the fact that the
% operator is commutative):

{\em
\begin{center}
\begin{tabular}{c@{\hspace{3em}}c}
\begin{tabular}{c}
$\transits{q_1}{a}{q'_1}$, $a\notin{\alpha}M_2$\\
\hline
$\transits{(q_1, q_2)}{a}{(q_1', q_2)}$\\
\end{tabular} 
& 

\begin{tabular}{c}
$\transits{q_2}{a}{q'_2}$, $a\notin{\alpha}M_1$\\
\hline
$\transits{(q_1, q_2)}{a}{(q_1, q_2')}$\\
\end{tabular} 
\end{tabular}
\end{center}
}

{\em
\begin{center}
\begin{tabular}{c} 
$\transits{q_1}{a}{q_1'}$,
$\transits{q_2}{a}{q_2'}$,
 $a \neq \tau$\\ \hline
$\transits{(q_1, q_2)}{a}{(q_1', q_2')}$\\
\end{tabular}
\end{center}
}

%\begin{note} ~\\
%\label{note:parcomp}
The language of $M_1 \parallel M_2$ is $\lang{M_1 \parallel M_2} = \{ t \mid
\project{t}{\alpha M_1} \in \lang{M_1} \land \project{t}{\alpha M_2}
\in \lang{M_2} \land t \in (\alpha M_1 \cup \alpha M_2)^\ast \}$
%\end{note}

%\cnote{add some properties here: transitivity}

\subsubsection{Properties} 
In this paper, we address assume-guarantee reasoning in the
context of checking safety properties.  %For liveness properties please look at~\cite{liveness}.  
For the context of our presentation, a
safety property is modeled as a deterministic CSM $P$, whose 
language $\lang{P}$ defines the set of acceptable
behaviors over $\alpha P$.  For CSMs $M$ and $P$ where $\alpha{}P
\subseteq \alpha{}M$, $M \models P$ if and only if 

\[\forall t \in \lang{M} : \project{t}{\alpha P} \in \lang{P}.\]

Checking properties reduces to reachability checks: 
$M \models P$ holds iff a special error state is unreachable in 
$M \parallel P_{err}$, where $P_{err}$ is the complement of $P$ and is obtained by completing $P$ such
that each missing transition on $\alpha P$ becomes a transition 
to the error state.

\subsection{Assume-guarantee Reasoning}

In the assume-guarantee paradigm a formula is a triple $\triple{A}{M}{P}$, 
where $M$ is a component, $P$ is a property, and
$A$ is an assumption about $M$'s environment.  The formula is true if 
whenever $M$ is part of a system satisfying $A$, then the system must
also guarantee $P$, \ie, $\forall E$, 
$E \parallel M\models A$ implies  $E \parallel M\models P$. 
Note that when $\alpha P \subseteq \alpha A \cup
\alpha M$, this is equivalent to $A \parallel M \models P$.

Let $M$ be a finite-state component with $\Sigma$ being the set of its
interaction points with the environment, i.e. the set of actions which
will participate in the composition of $M$ with another component from
the environment.  Furthermore, let $P$ be a safety property. Then
there is a natural notion of the {\em weakest assumption} $A_w$ for
$M$ with respect to $P$, with $\alpha A_w = \Sigma$. $A_w$ is
characterized by two properties:
\begin{itemize}
\item{\em Safety:} $\triple{A_w}{M}{P}$ holds.  
\item{\em Permisiveness:} $A_w$ characterizes all the possible
environments $E$ under which $P$ holds, \ie $\forall E: M\parallel 
E\models P \Rightarrow E\models A_w$.
\end{itemize}

These two conditions essentially ensure that $\forall E: M\parallel 
E\models P$ iff $E\models A_w$. It has been shown that, 
for any finite-state component $M$, the
weakest assumption $A_w$ exists, and can be constructed
algorithmically.  The weakest assumption is
associated with a similar notion of precision defined in the literature for
``temporal'' component interfaces~\cite{DBLP:conf/sigsoft/HenzingerJM05}, \ie, interfaces
that capture ordering relationships between invocations of component
methods.  For example, an interface may describe the fact that closing
a file before opening it is undesirable because an exception will be
thrown.  An ideal interface should precisely represent the component
in all its intended usages.  It should be {\em safe}, meaning that it
should exclude all problematic interactions, and {\em permissive}, in
that it should include all the good interactions~\cite{DBLP:conf/sigsoft/HenzingerJM05}.

%Safety and permissiveness can similarly be defined for assumptions,
%where the weakest assumption is the one that is both safe and
%permissive.  An assumption $A$ is safe, if $\triple{A}{M_1}{P}$. 

Similarly, assumption safety is concerned with restricting behaviors to only those that
satisfy $P$.  Permissiveness is concerned with including behaviors,
making sure that behaviors are restricted only if
necessary. Permissiveness is desirable, because $A_w$ is then
appropriate for deciding whether an environment $E$ is suitable for
$M_1$ (if $E$ does not satisfy $A_w$, then $E \parallel M_1$ does not
satisfy $P$). 

The simplest assume-guarantee rule is for checking a safety property
$P$ on a system with two components $M_1$ and $M_2$.

\paragraph*{Rule {\sc ASym}}

\[
\begin{array}{ll}
1: & \triple{A}{M_1}{P}\\
2: & \triple{\emph{true}}{M_2}{A}\\
\hline
   & \triple{\emph{true}}{M_1 \parallel M_2}{P}
\end{array}
\]

In this rule, $A$ denotes an assumption about the environment of
$M_1$. Note that the rule is not symmetric in its use of the two components,
and does not support circular reasoning. Despite its simplicity, experience
has shown it to be quite useful in the context of checking safety properties.

\subsubsection{Soundness and Completeness}

Soundness of an assume-guarantee rule means that whenever its premises
hold, its conclusion holds as well. Without soundness, we cannot rely
on the correctness of conclusions reached by applications of the rule,
making the rule useless for verification.  
Completeness states that whenever the conclusion of the rule is
correct, the rule is applicable, i.e., there exist suitable
assumptions such that the premises of the rule hold. 
Completeness is not needed to ensure correctness, but
it is an important measure for the usability of the rule.
Rule {\sc ASym} is both sound and complete. To show soundness, 
note that $\triple{\emph{true}}{M_2}{A}$ implies
$\triple{\emph{true}}{M_1\parallel M_2}{A}$. 
Then, since $\triple{A}{M_1}{P}$
also holds, it follows that $\triple{\emph{true}}{M_1\parallel M_2}{P}$
holds as well (from the definition of
assume-guarantee triples). 
Completeness holds trivially, by substituting $M_2$ for $A$.

For the use of rule {\sc ASym} to be justified, the assumption should
be (much) smaller than $M_2$, but still reflect $M_2$'s behavior,
i.e. $A$ should be an abstraction of $M_2$, according to
premise~2. Additionally, an appropriate assumption for the rule needs
to ``restrict'' $M_1$ enough to satisfy $P$ in premise~1. Coming up
with such assumptions manually is highly non-trivial.  In
the next sections we describe techniques for synthesizing
assumptions automatically.

\subsection{Abstraction}

To check properties of infinite-state components, we build {\em may} and 
{\em must} abstractions of software components. A popular technique uses 
predicate abstraction -- a special instance of abstract
interpretation~\cite{DBLP:conf/popl/CousotC77} that maps a potentially
infinite state transition system into a finite state transition
system via a finite set of predicates  $\Preds = \{p_1, \dots,
p_n\}$ over a program's variables. 

An abstraction function $\abst$ maps a concrete state $q$ to a set of
predicates that hold in $q$: $\abst(q) = \{p \in \Preds\ |\ q \models
p\}$. For a concrete transition we define corresponding {\em may} and
{\em must} transitions. Let $q_A, q'_A$ denote abstract states, and $q, q'$ denote concrete states (in the un-abstracted system):

\begin{itemize}
\item $q_A \stackrel{a}{\longrightarrow}_{must}q'_A$ iff $\forall q$
  s.t. $\abst(q) = q_A$, 
  $\exists q'$ s.t. $\abst(q') = q'_A$ and $q
  \stackrel{a}{\longrightarrow}q'$. 
\item $q_A \stackrel{a}{\longrightarrow}_{may}q'_A$ iff 
  $\exists q$ s.t. $\abst(q) = q_A$ and 
  $\exists q'$ such that $\abst(q') = q'_A$ and $q
  \stackrel{a}{\longrightarrow}q'$. 
\end{itemize}

Given component with communicating state machine $C$, the must and may
abstractions with respect to the set of abstract predicates $\Preds$
are defined as
$C^{{must}}_\Preds=(2^{\Preds},\Sigma,\abst(q_0),\longrightarrow_{must})$
and $C^{{may}}_\Preds=(2^{\Preds}, \Sigma, \abst(q_0),
\longrightarrow_{may})$, respectively. 
We write $C^{must}$ or $C^{may}$ when $\Preds$ is clear from the context.

The {\em must} abstraction consists of behaviors
which are guaranteed to be present in the concrete (un-abstracted)
component; it represents an under-approximation as it
might miss some concrete behaviors.  The {\em may} abstraction
represents an over-approximation; it consists of all
concrete behaviors of the concrete component but it may also
contain additional, spurious ones.

Algorithms for computing may and must abstractions with the help of a
theorem prover are given in e.g.~\cite{DBLP:conf/padl/PodelskiR07}.
For automated abstraction refinement, we use weakest precondition
calculations over counterexample
traces~\cite{DBLP:conf/cav/ClarkeGJLV00,DBLP:conf/cav/NamjoshiK00}. Let
$\phi$ be a predicate characterizing a set of states. The weakest
precondition of $\phi$ with respect to a transition $\tau_i$ is
$wp(\phi,\tau_i)=\{ q | (q\stackrel{\tau_i}{\longrightarrow}q'
\Longrightarrow \phi(q'))\}$ and it characterizes the largest set of states
whose successors by transition $\tau_i$ satisfy $\phi$.

%{\bf expand: explain that refinement consists of adding more abstraction predicates, discovred using %simulation of counterexamples and weakest precondition calculations.}

%Note that the set of may transitions is a super-set of the must
%transitions. % and one can build the must abstractions by simply removing
%the may transitions (i.e. the transitions that are not in the must
%abstractions).
From the above definitions it follows that
the may and must abstractions define simulations
between $C^{must}$ and $C$, and between $C$ and
$C^{may}$, respectively. Since simulation implies trace
inclusion, we have the following characterization of under- and over-
approximations:

\[\lang{C^{must}}\subseteq \lang{C} \subseteq \lang{C^{may}}\]

\subsection{The L* Algorithm}

L* was developed by Angluin~\cite{angluin87} and later improved by
Rivest and Schapire~\cite{rivest93}. L* learns an unknown regular
language $U$ over alphabet $\Sigma$ and produces a {\em minimal
  deterministic} finite state automaton (DFA) that accepts it.  L*
needs to interact with an oracle, called a \emph{Minimally Adequate
  Teacher}, that answers two types of questions from L*. The first
type is a \emph{membership query} asking whether a string $\sigma \in
\Sigma^*$ is in $U$. For the second type, the learning algorithm
generates a \emph{conjecture} $A$ and asks whether $L(A) = U$. If
$L(A) \neq U$ the Teacher returns a counterexample, which is a string
$\sigma$ in the symmetric difference of $L(A)$ and $U$.  L* is
guaranteed to terminate with a minimal automaton $A$ for $U$.  If $A$
has $n$ states, L* makes at most $n-1$ incorrect conjectures.  The
number of membership queries made by L* is $O(kn^2 + n \log m)$, where
$k$ is the size of $\Sigma$, $n$ is the number of states in the
minimal DFA for $U$, and $m$ is the length of the longest
counterexample returned when a conjecture is made.

%{\bf give examples of finite systems, properties, assumptions: input-output. Gove also example for %abstrcation}

\section{Compositional Verification for Finite-State Systems}

\subsection{Learning Assumptions}

From the definition of the
weakest assumption $A_w$, one can observe
that for $A_w$, the premises of Rule {\sc ASym}  
become necessary, in addition to being sufficient, for the
conclusion of the rule to hold. In other words,
($\triple{A_w}{M_1}{P}$) and ($\triple{\emph{true}}{M_2}{A}_w$)
hold, if and only if  
($\triple{\emph{true}}{M_1 \parallel M_2}{P}$).
This is an advantage for an automated assume-guarantee 
reasoning framework, since it enables us to also
disprove properties of a system, compositionally.

\begin{figure}
\includegraphics[scale=0.30]{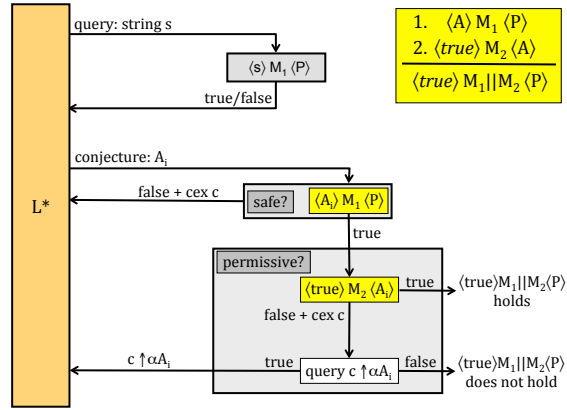}
\centering
\caption{Learning Assumptions for Assume-Guarantee Reasoning}
\label{fig:AGframework} %\vspace*{-.3cm}
\end{figure}

The framework illustrated in Figure~\ref{fig:AGframework}, 
and first presented in~\cite{cobleigh03}, provides a learning-based 
approach to assume-guarantee reasoning of finite-state components. 
In this framework, L* targets the computation of
the weakest assumption $A_w$, and its application to rule {\sc ASym}. 
The set of communicating actions of component $M_1$ with its environment
is defined as:
$(\alpha M_1 \cup \alpha P) \cap \alpha M_2$.
We use this set as the alphabet of the weakest assumption in this context,
hence the alphabet over which L* is learning.
Note that the framework uses the knowledge of the actual
environment of component $M_1$, namely, component $M_2$, to
make the reasoning more efficient. More specifically, the framework 
implements a teacher for L*, 
meaning that it responds to queries and conjectures, 
as described in the following.

\noindent
\emph{Queries.}
L* is first used to repeatedly {\em query} $M_1$ to check whether, 
in the context of strings $s$, $M_1$ violates the property. 
More formally, the query corresponds to checking the 
triple $\triple{s}{M_1}{P}$
as illustrated in Figure~\ref{fig:AGframework}. 
Checking $\triple{s}{M_1}{P}$
corresponds to simulating string $s$ on $M_1 \parallel P$: if 
an error is reachable, then the triple is \emph{false}, otherwise 
it is \emph{true}.
The query returns \emph{true}/\emph{false}
if $\triple{s}{M_1}{P}$ is \emph{true}/\emph{false}, respectively.
This is because, as mentioned, 
$A_w$ allows all behaviors that satisfy the property, and disallows
only violating behaviors. 

\noindent
\emph{Conjectures.}
The automaton $A_i$ conjectured during iteration $i$, 
is checked for correctness, which
in this context means checking whether it corresponds to the weakest 
assumption or not. 
As discussed earlier, the weakest assumption is safe and permissive.
We therefore reduce equivalence queries to two separate checks,
for safety, and permissiveness of $A_i$.

The first check is for safety: $\triple{A_i}{M_1}{P}$; it 
can be performed by a model checker. 
If $A_i$ is safe, then the
teacher proceeds to checking permissiveness. If it is unsafe, the model checker
returns a counterexample. The resulting counterexample $c$,
projected on the assumption alphabet $\alpha A_i=\alpha A_w$, 
is returned to L* to
refine its conjecture. The projection is necessary because L* needs
counterexamples in terms of the alphabet over which it is learning.

As discussed, permissiveness is concerned with ensuring
that the assumption does not exclude correct behaviors.
However, given the fact that the main goal of the framework
is  to
prove or disprove a property on the system using
assume-guarantee reasoning, the framework does not need
to generate a fully permissive assumption. Rather, it uses $M_2$
to add behaviors to over-restrictive assumptions on demand, and 
as needed for completion of the verification. 

Note that the check for safety coincides with premise 1 of Rule {\sc ASym}.
It therefore remains
to check premise 2 ($\triple{\emph{true}}{M_2}{A_i}$).  
We use premise 2 to drive the permissiveness check
and potentially complete assume-guarantee reasoning 
in Figure~\ref{fig:AGframework} as follows.
If $\triple{\emph{true}}{M_2}{A_i}$ 
is \emph{true}, then we know that both
premises of Rule {\sc ASym} hold, and therefore
that $P$ holds for $M_1 \parallel M_2$.
If $\triple{\emph{true}}{M_2}{A_i}$ is \emph{false},
the Teacher performs some analysis
to determine the underlying reason (see Figure~\ref{fig:AGframework}).
The Teacher performs a query (of the L* type) in order to determine 
whether the returned counterexample $c$, 
projected to the alphabet of the assumption,
belongs to $A_w$, in which case L* needs to refine the assumption.
If the query returns \emph{true}, then $A_i$ is not permissive, 
so  $\project{c}{\alpha A_i}$
is returned to L* for refinement of its guess. If, on the other hand, the 
answer is false, it means that $c$ is a word that belongs to $M_2$,
in the context of which $M_1$ violates the property $P$. As a consequence,
$M_1 \parallel M_2$ does not satisfy the property $P$.

Each new assumption marks the beginning of the next iteration
cycle. 
Notice that the answers that the framework provides to
L* are always precise with respect to the targeted weakest assumption. 
However, 
the framework uses $M_2$ to select which missing words to 
include in the language of the assumption.
The reason is that we restrict our reasoning to a specific context,
rather than accounting for all possible contexts, as required
for the computation of $A_w$. 
That means, of course, that the assumption obtained from this framework  
does not necessarily correspond to $A_w$. 
On the other hand, we remind the reader that the primary goal is to obtain conclusive results
from the assume guarantee rule. As soon as we are able to prove or disprove the
property in the system, we stop refining the learned assumption, since we 
have achieved our goal.
The assumption computed with this framework will be smaller than, or in the worst case equal to $A_w$ in terms of number of states, 
as 
guaranteed by the characteristics of L*.  In the worst case,  where $A_w$ itself is computed, the framework is guaranteed
to terminate, because $A_w$ is both necessary and sufficient, and therefore the framework will prove or 
disprove the property during this iteration. 

\subsection{Correctness Arguments.}

{\em Framework correctness argument}:
The framework directly uses
the assume-guarantee rule Rule {\sc ASym} to answer conjectures.
Soundness of the rule guarantees  correctness of the positive answers
by the framework. 
On the other hand, each counterexample reported is a real
counterexample, as discussed above.

{\em Teacher correctness argument}:
Correctness of the teacher corresponds to showing that all
the answers returned to L* are consistent with $A_w$. This was 
discussed during the presentation of the framework above.

{\em Termination argument}:
Since the Teacher implemented in our framework
only comes back to L* for refinement with counterexamples 
related to $A_w$,
the framework eventually converges to $A_w$,
unless it terminates earlier.
As discussed, $A_w$ makes Rule {\sc ASym} sound and complete,
and therefore our framework will return a conclusive answer at that iteration.

\subsection{Extensions}

The framework has been extended to reasoning about more than two
components (by applying the framework recursively for Premise 2) and
to other circular and symmetric
rules~\cite{DBLP:journals/fmsd/PasareanuGBCB08}. It has been
demonstrated on checking flight software models, where it achieved
significant savings in terms of time and memory, and in some cases it
was able to terminate while the monolithic (non-compositional)
verification ran out of time and memory resources.

\section{Compositional Verification for Infinite-State Systems}
\label{sec:compverif}

\begin{figure}
\includegraphics[scale=0.50]{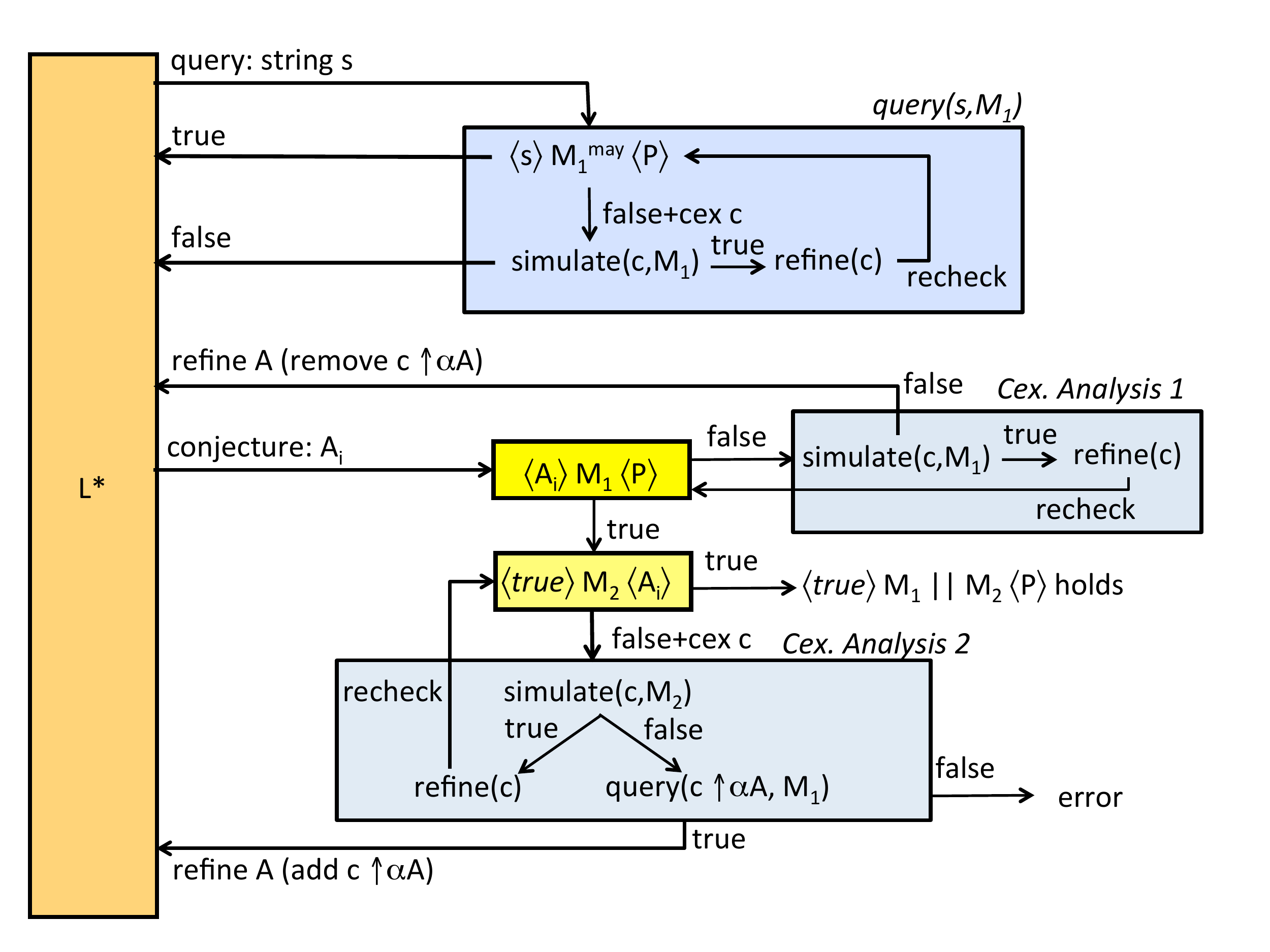}
\centering
\caption{Automated Compositional Verification for Infinite-State Systems}
\label{fig:AGInfinite} %\vspace*{-.3cm}
\end{figure}

When reasoning about infinite-state components, abstractions need to
be used to further reduce the state spaces of individual components.
Figure~\ref{fig:AGInfinite} illustrates a learning-based framework for
automating compositional verification for infinite-state systems.

In this section, we will first present how assume-guarantee reasoning
of infinite-state components can be performed given some assumption $A$.
We will subsequently discuss how such steps are introduced in creating 
automated assume-guarantee frameworks using learning for assumption inference.
 
\[
\begin{array}{ll}
1: & \triple{A}{M_1}{P}\\
2: & \triple{\emph{true}}{M_2}{A}\\
\hline
   & \triple{\emph{true}}{M_1 \parallel M_2}{P}
\end{array}
\]

Going back to rule {\sc ASym}, let us assume
that both $M_1$ and $M_2$ are infinite-state, or too large to perform
the steps involved in the two premises by model checking. A standard
approach used in such cases is to use abstractions of components $M_1$ and
$M_2$. 

As discussed in Section~\ref{sec:formalisms}, finite
over-approximations, or \emph{may} abstractions, of a component, are
typically used to guarantee correctness with respect to some
property. May abstractions have more behavior than the system that
they abstract. As such, counterexamples that are detected using may
abstractions may be spurious. On the other hand, finite
under-approximations, or \emph{must} abstractions, are better suited
for detecting property violations that are not spurious.
Assume-guarantee rules allow us to infer correctness of a system based
on local correctness checks of the system components. In this setting,
it makes sense to use may abstractions of components $M_1$ and $M_2$,
since $\triple{A}{M^{may}_1}{P} \Longrightarrow \triple{A}{M_1}{P}$
and $\triple{\emph{true}}{M^{may}_2}{A} \Longrightarrow
\triple{\emph{true}}{M_2}{A}$.

We can therefore use the {\em may} abstractions to check the two
premises of the assume-guarantee rule. If both premises hold for the
may abstraction it follows that the premises hold for the original
un-abstracted components, and hence $P$ holds on the composition $M_1
\parallel M_2$.  Let us now analyze the two premises given $M^{may}_1$
and $M^{may}_2$.

\begin{itemize}
\item{\em Premise 1:} If $\triple{A}{M^{may}_1}{P}$ holds, then we
  know that $\triple{A}{M_1}{P}$ holds, which completes this check. If
  $\triple{A}{M^{may}_1}{P}$ does not hold, then we obtain a
  counterexample, say $c$.  This counterexample may be spurious either
  due to abstraction or due to the approximation introduced by
  assumption inference.  Similarly to a CEGAR approach, we can
  determine whether $c$ is spurious due to abstraction by checking if
  it can be simulated to completion on the infinite component $M_1$,
  using an operation $simulate(c,M_1)$. Since $c$ is finite,
  simulation is possible. If $c$ is not a valid execution of $M_1$, it
  means that it is a spurious counterexample, due to abstraction. If
  it is a valid execution of $M_1$, then it means that premise 1 does
  not hold, and therefore $A$ is too approximate to make premise 1
  pass. Below we will describe a new learning framework that is able to use 
  this information to
  automatically refine the abstraction $M^{may}_1$ or the assumption
  $A$, respectively. 
%We will explain the specific refinement steps in the
 % context of the learning framework below.

\item{\em Premise 2:} If $\triple{\emph{true}}{M^{may}_2}{A}$ holds,
  then we know that \triple{\emph{true}}{M_2}{A} holds, which
  completes this check. If $\triple{\emph{true}}{M^{may}_2}{A}$ does
  not hold, then we obtain a counterexample, say $c$, which needs to
  be analyzed.  This time, $c$ may be spurious due to the abstraction
  of $M_1$ {\em or} $M_2$; furthermore, $c$ may be spurious due to the
  approximation in the assumption. We therefore simulate $c$ on $M_2$,
  using operation $simulate(c,M_2)$. If $c$ is not a valid execution
  of $M_2$, it means that it is a spurious due to
  abstraction in $M_2$, and abstraction-refinement is needed for $M_2$. 
  If $c$ is a
  valid execution of $M_2$, then it means that premise 2 does not
  hold, and this may be an indication of a real error or it may mean
  that the counterexample is spurious either due to abstraction of
  $M_1$ or to approximation in $A$.  Below, we describe how the
proposed learning
  framework uses this information to automatically refine the
  component abstractions or the assumption $A$, as needed.
\end{itemize}

\subsection{Learning for Assumption Inference}

%{\bf explain what $\alpha A$ is; recap that we try to leanr the weakest assumption; other flavors exists; even the simulation needs more explanation; is it simply based on alphabets or more than that, i.e. does it need to consider states? need to explain refinement formally}

Let us now revisit the learning framework for finite-state systems and
see how it can be extended for infinite-state. As discussed, L* needs
a teacher that can answer membership queries and conjectures. 

\begin{itemize}
\item{\em Membership queries:} A membership query needs to check if a
  finite word $s$ over the alphabet $\alpha A_w$ should be included in
  the language of $A_w$.  Similar to the finite-state case (see
  Figure~\ref{fig:AGframework}) we are interested in finding out
  whether in the context of $s$, $M_1$ violates $P$ or not.  
  We first check $\triple{s}{M_1^{may}}{P}$. If it leads to no error, it means this is true for $M_1$ as well (since the abstraction is conservative) and {\em true} is returned to L*. If an error is detected, the reported counterexample $c$ is checked to see if it corresponds to a real trace in $M_1$.
This
  amounts to symbolically simulating $c$ on $M_1$, denoted by 
  $simulate(c,M_1)$. Since $s$ is finite, $c$ is finite too so the
  simulation is possible and will terminate. If the result of simulation is that this is a real trace, indicating a real error in $M_1$, {\em
    false} is returned to L*. However, if this is not a real error,   the spurious counterexample {\em c} is used to refine $M_1^{may}$
      and the membership check is repeated. The abstraction-refinement
      denoted by $refine(c)$ is described in more detail below.
\end{itemize}

When L* produces an assumption $A_i$, then the new framework needs to
check whether $A_i$ can be used to complete the assume-guarantee
reasoning of the system.  In other words, it needs to check Premises 1
and 2. For these checks, the framework uses $M^{may}_1$ and
$M^{may}_2$, respectively, as described above.

When the safety check associated with premise 1 passes, then the
framework proceeds with the permissiveness check involving $M_2$.  
If, however, a
counterexample $c$ is obtained, then $c$ needs to be analyzed as
described below (see Figure~\ref{fig:AGInfinite}).

\begin{itemize}
\item{\em Counterexample Analysis 1:} If the result of $simulate(c,M_1)$
is that $c$ is a violating 
execution of $M_1$ then $A_i$ must be refined; $\project{c}{\alpha A_i}$ is returned to L*,
and L* will work on creating a new approximation of the assumption. If $c$ is not
violating in $M_1$ then $M^{may}_1$ must be refined, using $refine(c)$
and the safety check is performed again with the new abstraction.
\end{itemize}

The permissiveness check consists of applying premise 2 using $M^{may}_2$ 
as described above. If the check passes, we can conclude that 
 $\triple{\emph{true}}{M_1 \parallel M_2}{P}$ holds. If 
not, we analyze the returned counterexample
$c$ as described below (see Figure~\ref{fig:AGInfinite}). 

\begin{itemize}
\item{\em Counterexample Analysis 2:} If the result of $simulate(c,M_2)$ shows that $c$ is not a real counterexample, then $M^{may}_2$ is
  refined using $refine(c)$, and the permissiveness check is repeated.
  If the result however indicates a real trace in $M_2$, then we must further analyze
  $c$ to determine if it uncovers a real violation in the system or
  not.%, similarly to the case of the finite-state framework. 

\Comment{We first
  check if $simulate(\project{c}{\alpha A_w},M_1)$ indicates a real violation
  in $P$.  If it does, this means that $c$ exposes a real error in
  $M_1 \parallel M_2$, which is reported to the user.

Otherwise, it may be the case that either $M_1^{may}$ or $A_i$ need to
be refined. 
}
Similar to the finite-state case (Figure~\ref{fig:AGframework}), 
this further analysis amounts to performing a query for $\project{c}{\alpha A_w}$ on $M_1$. 
Note that in the infinite-state case, performing a query might result in further refinements for $M_1$ and corresponding subsequent checks for the new abstractions (as described in the query check above). If the result of the analysis is that the assumption needs to be refined, the counterexample $\project{c}{\alpha A_w}$ is returned to L*, which will work on creating a new approximation for $A$, based on new membership queries and conjecture checks.
\end{itemize}

Note that this framework has an important characteristic: the
information that is communicated to L* is always correct with respect
to the concrete system.  As a result, refinement of abstractions does
not require for L* to restart learning.  

In our proposed framework, abstraction refinement is applied whenever
a violating trace $t$ is discovered that belongs to a {\em may}
abstraction ($M_1^{may}$ or $M_2^{may}$) but not to the corresponding
un-abstracted components ($M_1$ and $M_2$ respectively). Consequently
$t$ must contain a {\em may} transition that has no correspondence in
the un-abstracted system. We use a simple strategy based
on weakest precondition calculations~\cite{cav2010} to compute new abstraction
predicates that are guaranteed to eliminate the spurious
transitions. Given a counterexample $t$ as a sequence of transitions
$\{\tau_1,\tau_2 ...\tau_n\}$ we compute refinement predicates $wp(true,t)$
by using
weakest preconditions recursively based on the following definition
$wp(\phi, t)=wp(wp(\phi,\tau_n),\{\tau_1,\tau_2 ...\tau_{n-1}\})$.

%\subsection{Abstraction for Assumption Inference}

%aka AGAR

%{\bf CUT this part??}

\subsection{Correctness and Termination}

%****** Can be simplified ...

%The argument should hold for both.

We argue now the correctness and the termination for the proposed
algorithms.  Our framework returns true only if both
$\triple{A}{M_1^{may}}{P}$ and $\triple{\emph{true}}{M_2^{may}}{A}$
hold. Since the assume guarantee rule is sound, it follows that
$\triple{\emph{true}}{M_1^{may}\parallel M_2^{may}}{P}$ holds, and
since the {\em may} abstractions are over-approximations, it follows
that $\triple{\emph{true}}{M_1^{may}\parallel M_2^{may}}{P}$ holds as
well. On the other hand, if the framework reports an error, it finds a
trace which is both a trace of $M_2$ and of $M_1$ and it
leads to a violation of $P$, hence it is an error in $M_1 \parallel
M_2$ as well.  

%We therefore have the following correctness result.

%\begin{theorem}[Correctness]
%If our proposed framework terminates, then it returns {\em true} if
%$P$ holds in $M_1 \parallel M_2$ and {\em false} otherwise.
%\end{theorem}

For infinite-state components, the abstraction-refinement algorithm
used in our proposed framework may not always terminate. However, from
previous work on automatic abstraction
refinement~\cite{DBLP:conf/cav/NamjoshiK00}, we know that if a
component $M$ has a finite bisimulation quotient, then
abstraction-refinement (based on weakest preconditions calculations)
converges to that finite quotient. It follows that there is a
refinement iteration bound $i$ such that $M^{may}$ is bisimilar to $M$
(in our case the argument applies to $M_1$ and $M_2$).  Since
bisimulation implies trace equivalence it follows that
$\lang{M^{may}}=\lang{M}$ at that bound. Once that bound is reached,
no abstraction-refinement is performed, and the obtained
counter-examples will either be returned to L* for
assumption-refinement or they will be returned to the user as real
errors, in which case the computation will terminate.  By the
correctness of L* we are guaranteed that it will eventually produce
$A_w$ wrt. $M_1^{may}$ and property $P$. During this step, checking
premise 1 will return true (by definition of $A_w$) and checking
premise 2 will either return true and terminate, or return a
counterexample. This counterexample is a trace in
$\lang{M_2^{may}}=\lang{M_2}$ that is not in $\lang{A_w}$. Since $A_w$
is both safe and permissive, counterexample analysis will return
false and the framework will terminate.

%We therefore have the following termination result.

%\begin{theorem}[Termination]
%If both $M_1$ and $M_2$ have finite bisimulation quotients, then the framework 
%terminates.
%\end{theorem}

The framework may terminate earlier, not necessarily when
it reaches bisimulation quotients for $M_1$ and $M_2$, but as soon as
the abstractions $M_1^{may}$ and $M_2^{may}$ and the inferred
assumption $A$ are good enough to show that the two premises hold or to
expose a real counterexample.

\section{Example: A simple communication protocol}

In this section, we illustrate the presented algorithms through a very simple communication protocol. The protocol consists of two components, \emph{Sender} and \emph{Receiver}. We analyze two different versions of the protocol, one for the case where its components are finite-state, and one where they are infinite-state.

\subsection{Finite State Protocol}
\begin{figure}
\includegraphics[scale=0.30]{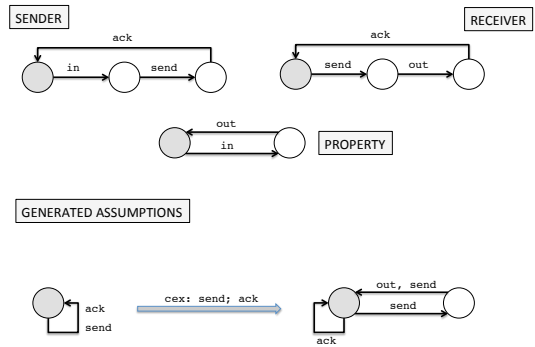}
\centering
\caption{Learning-based AG reasoning for a finite system}
\label{fig:finout} \vspace*{-.5cm}
\end{figure}

The \emph{Sender} and \emph{Receiver} for the finite-state case are illustrated in Figure~\ref{fig:finout}. The \emph{Sender} starts by receiving input from the environment. It subsequently sends a message to the \emph{Receiver}, and waits for an acknowledgement. The \emph{Receiver}, upon a message being sent by the \emph{Sender}  produces an output, and subsequently acknowledges receipt of the message. The desired property from the environment's perspective is that actions \emph{in} and \emph{out} alternate, with \emph{in}  occurring first. As illustrated, the compositional verification framework produces two assumptions; the second assumption is suitable for completing the assume-guarantee reasoning for this example. Note that the assumption has 2 states, which is one state less than the \emph{Receiver}. The example is small and used for illustrative purposes, so the benefits of compositional verification are modest. 

\subsection{Infinite State Protocol}

\begin{figure}
\includegraphics[scale=0.30]{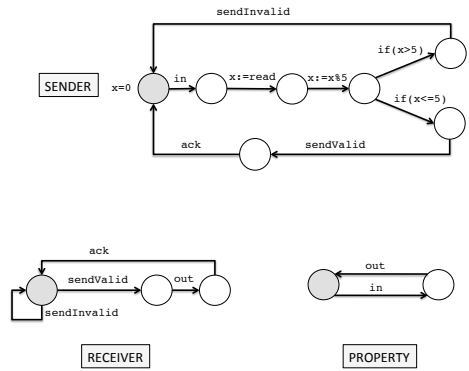}
\centering
\caption{Infinite state communication protocol}
\label{fig:infinout}  \vspace*{-.5cm}
\end{figure}

Now let us modify the example with an infinite-state \emph{Sender}. The \emph{Sender} has a variable $x$, initially set to $0$, and where $x$ is in the domain of natural numbers. In Figure~\ref{fig:infinout}, we represent the component in terms of pseudo-code, displayed in terms of its control flow. Similarly to the finite-state case, the \emph{Sender} starts by receiving input from the environment, which prompts it to perform a \emph{read}, the results of which are stored in $x$. In other words, this represents the non-deterministic assignment of a natural number value to variable $x$, modeling in term the communication of a value from the environment. Subsequently, $x$ is set to its previous value modulo $5$. The control flow then branches depending on whether $x > 5$, in which case it communicates an invalid message, and otherwise it sends a valid message, and waits for an acknowledgement from the \emph{Receiver}.  The \emph{Receiver} may now receive a valid message, in which case it behaves as in the finite-state case, but also an invalid message. In the latter case, it directly transits to its initial state and waits for the next message. The property is identical to the finite-state case.
  
\begin{figure}
\includegraphics[scale=0.30]{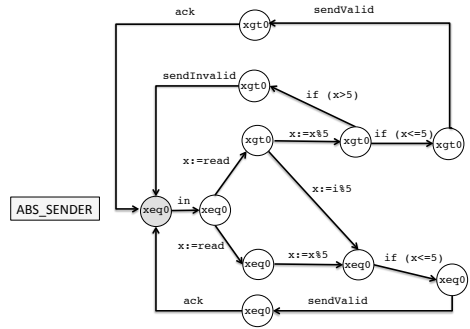}
\centering
\caption{Abstracted sender - version 1}
\label{fig:abstractedIn} \vspace*{-.5cm}
\end{figure}

\begin{figure}
\includegraphics[scale=0.30]{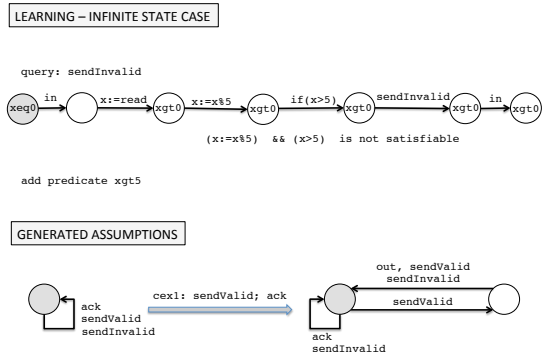}
\centering
\caption{Learning-based AG reasoning for infinite system}
\label{fig:learningInOut} \vspace*{-.5cm}
\end{figure}

Assume that one first creates an abstraction of \emph{Sender} based on predicates $x = 0$ and $x > 0$. The abstraction is illustrated in Figure~\ref{fig:abstractedIn} whereas, Figure~\ref{fig:learningInOut} depicts the learning process for this infinite state system. As illustrated in Figure~\ref{fig:learningInOut} , when the learning algorithm queries the abstraction for \emph{``sendInvalid"}, the answer to the query is negative. The detailed trace at the top of Figure~\ref{fig:learningInOut} provides the exact trace of the abstracted sender that violates property $P$. However, if this trace is simulated on the concrete \emph{Sender}  component of Figure~\ref{fig:infinout}, we can detect that it is impossible for $(x>5)$ to hold right after performing operation $(x:=x\%5)$; therefore, this initial abstraction needs to be refined. If one adds predicates $x >5$ and $x \leq 5$, then the learning framework is able to show that $P$ is satisfied with assume guarantee reasoning. Note that the assumptions obtained are similar to the infinite-state case, where $send$ is now split in  two actions, $sendValid$ and $sendInvalid$. 

\section{Interface Generation for Infinite-State Components}
\label{sec:interface}

Assumptions are closely related to the notion of 
component interfaces. Intuitively, a component interface
summarizes aspects of a component that
are relevant to its customers. 
Traditionally, component interfaces
have been of a purely syntactic form, that included information
about the services/methods that can be invoked on the component,
and their signatures, meaning the numbers and types of arguments and
their return values. However, there is a recognized need for richer interfaces that capture
additional, behavioral, aspects of a component. 
%With the advent of component-based and
%distributed development,
%service-oriented computing, and other such concepts,
%components are no longer viewed as parts of specific systems,
%but rather as open systems that can be reused, or connected
%dynamically, in a variety of environments to form larger systems.
%However, components may also not publish their internal details.
%Interfaces must therefore step up to the role of representing
%component aspects that are relevant for various tasks such as 
%dynamic component retrieval and substitution, or functional
%and non-functional reasoning.

Temporal interfaces, as introduced in Section~\ref{sec:formalisms}, 
are richer interfaces that capture ordering
relationships between invocations of component methods.
Ideally, a temporal interface should precisely represent the
component in all its intended usages.
In other words, it should include all the good
interactions, and exclude all problematic interactions. 

In such a context, component interfaces have
the same flavor as assumptions that relate to safety properties,
as studied in the previous section.
However typically an interface summarizes the component
{\em irrespective} of the environment in which 
the component is to be introduced, while 
we have seen that an assumption (used in compositional reasoning)
serves as a potentially
imprecise interface that is sufficient for breaking up
a targeted
verification problem into simpler problems;
all components that participate in the verification problem are
known and available.

In fact, a precise interface is similar to the weakest component 
assumption. It is therefore characterized in terms of two
properties, safety, and permissiveness. For simplicity, we consider here error states
that are not introduced by explicit properties, but
are rather assumed to represent undesirable component states 
(e.g. assertion violations in the component's code). 
Interfaces can be learned through frameworks similar to
those we developed for assume-guarantee reasoning; 
queries, and the part of conjectures related 
to safety, are answered in an identical way.
%, except that in the case of 
%interfaces we assume that error states exist in the components, whereas
%in the earlier frameworks they come from the property $P$.

\begin{figure}
\centering
\includegraphics[scale=0.30]{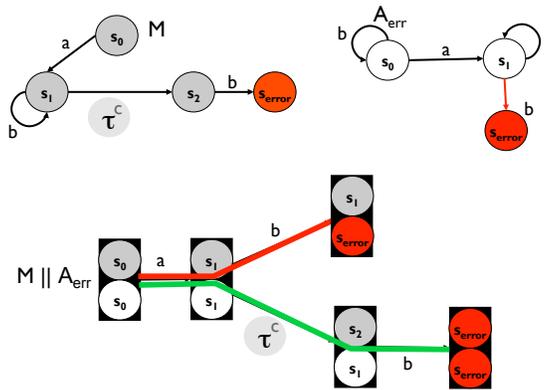}
\caption{Checking for Permissiveness}
\label{fig:permissiveness}
\end{figure}

Permissiveness, however, is more difficult because the environment of 
the targeted component is not available. 
The example in Figure~\ref{fig:permissiveness}  shows how a  permissiveness check could be performed.
Component $M$ has states named $s_0$, $s_1$, $s_2$, $s_{error}$ and interface
$A$ has states named $s_0$, $s_1$ and $s_{error}$. 
A permissiveness check needs to detect sequences that are blocked by the interface but legal 
in the component. Such sequences identify that the interface is not permissive. 
This can be performed by checking reachability, in $M \parallel A$, of states of the form $[s_i, s_{error}]$,
where $0 \leq i \leq 2$.
According to this check, trace $\langle a,b  \rangle$
leading to state $[s_1, s_{error}]$ in the composition could be an indication that $A$ is not permissive. 
However this is not true, since the same sequence of actions leads to $[s_{error}, s_{error}]$ on a different path, due to non-determinism. This happens because the alphabet of the assumption is $\{a, b\}$,
meaning that action $c$ in $M$ is considered as a $\tau$ from the point of view of $A$. 
In the figure, this is illustrated as a $\tau$ action covering action $c$.

This example illustrates the fact that non-determinism in component
$M$ may cause spurious counterexamples in the permissiveness
reachability check described above. As a consequence, precise
characterization of permissiveness requires determinization of
component $M$, which can be performed using subset
construction. The permissiveness check is therefore
NP-hard~\cite{DBLP:conf/popl/AlurCMN05}, and can be inefficient in practice.
Several approaches have been proposed to deal with this problem.
Unless determinization is a viable solution for a targeted component
$M$~\cite{DBLP:conf/cav/BeyerHS07}, heuristic approaches are often used to
determine whether a counterexample is spurious~\cite{DBLP:conf/popl/AlurCMN05,DBLP:conf/fase/GiannakopoulouP09}. 

Let us examine now the case of infinite-state component $M$. Abstraction is again needed to reason about such components. However using {\em may} abstractions alone turns out to be insufficient, because the generated interfaces may be overly restricting (due to the spurious error traces present in the abstraction).
In previous work~\cite{cav2010}, we have shown the following result:

\vspace{0.5cm}
{\em 
Assume a component $M$, a {\em may} abstraction $M^{may}$ and a {\em must} abstraction $M^{must}$ for $M$. If an interface $A$ for $M$ is {\em safe} with respect to $M^{may}$ and {\em permissive} with respect to $M^{must}$, then $A$ is safe and permissive with respect to $M$.
}
\vspace{0.5cm}

Based on this result, we developed a framework~\cite{cav2010} 
that interleaves abstraction-refinement and L* learning for the 
automated generation of interfaces for infinite-state components. 
The framework uses both $M^{may}$ and $M^{must}$ to compute an
interface for $M$ that is both safe and permissive. The abstractions
are refined automatically from counterexamples obtained during the
reachability checks performed by the framework.

It is interesting to note that in case component $M$
is {\em observationally} deterministic (i.e. deterministic with respect to its interface actions), the {\em must} abstraction is deterministic as well.
Thus its use enables us to avoid the exponentially expensive
determinization step that is required when working with non-deterministic components.
The idea of the framework, based on the result above, is to use
$M^{may}$ for the safety check, and $M^{must}$ for the permissiveness check.

The safety check is similar to the compositional verification
framework.
Let us therefore analyze how the permissiveness check is performed
with $M^{must}$, when $M$ is observationally deterministic. 
If in $M^{must} \parallel A$
it is not possible to reach an error state in $A$ that is an accepting
state in $M^{must}$, then it means that $A$ permissive, in which case
the framework produces $A$ as a safe and permissive interface for $M$.
If, however, such a combined state is reachable by some counterexample
$c$, then $c$ is analyzed as follows. If $c$ leads to an error state
in $M$, then $M^{must}$ needs to be refined to avoid this discrepancy.
Otherwise, $c$ represents a real counterexample to the permissiveness
of $A$ and it is returned to L* for refining the assumption.
If, on the other hand, $M^{must}$ is observationally non-deterministic, it must be determinized 
prior to performing the test.

In conclusion, we use both may and must abstractions for interface generation of infinite state components. 
In contrast, may abstractions are sufficient for compositional verification, because component   $M_2$ 
is known and can be used for selective permissiveness checks.

\Comment{
\dnote{shall we include a figure for this?}

\begin{theorem}[\cite{ourCAV}] 
Assume a component $C$, a {\em may} abstraction $C^{may}$ and a {\em must} abstraction $C^{must}$ for $C$. If an interface $A$ for $C$ is {\em safe} wrt $C^{may}$ and {\em permissive} wrt $C^{must}$, then $A$ is safe and permissive wrt $C$.
\end{theorem}

It follows that we can use $C^{may}$ and $C^{must}$ for computing an
interface for $C$ taht is both safe and permissive. The abstractions
are refined automatically from counterexamples obtained during the
reachability checks performed by our framework. In case component $C$
is deterministic, the {\em must} abstraction is deterministic as well,
and thus its use enables us to avoid an exponentially expensive
determinization step that is required when working with may
abstractions only.
}

\section{Related Approaches}
\label{sec:related}

%Sagar et al, our old work etc.

We briefly describe here some of the related work that combines abstraction and compositional reasoning in the context of infinite system analysis. 
The Magic tool  performs verification of concurrent,
message-passing C programs using abstraction-refinement in a compositional way; the work has been
extended with a two-level abstraction scheme, but it does not use assume-guarantee style verification. The Blast tool uses predicate abstraction and
has been extended to perform
assume-guarantee reasoning for checking race conditions in multi-threaded C
code \cite{blast}. In contrast to our work, it targets shared memory communicating programs, and therefore it uses a different style of assume guarantee rule. Moreover,
the approach used by Blast is not based on learning.
Bandera \cite{bandera, thesis} was aimed at verifying concurrent Java programs. 
Bandera employed data abstraction and modular reasoning with 
user-supplied assumptions, but not automated assumption generation and 
CEGAR, as we do here. As already mentioned, algorithms for interface synthesis for infinite state components have been presented in~\cite{DBLP:conf/popl/AlurCMN05,DBLP:conf/cav/BeyerHS07,cav2010}. Recently, L* learning has been combined with symbolic execution to automatically
generate interfaces for Java classes that include methods with parameters~\cite{DBLP:conf/sas/GiannakopoulouRR12}.

\section{Conclusion}
\label{sec:conclusion}

In this paper, we discussed the different types of abstractions that
can be used for applying learning-based assume-guarantee reasoning and
interface generation to infinite-state systems. We proposed a new
framework for automated compositional verification, and illustrated it
with a simple example. At the present time, we are not able to perform
extensive experimental evaluation of the framework. For the presented
example, we carried out all steps, except for abstraction refinement,
automatically. In the future, we plan to implement and evaluate
extensively the proposed framework. We are also interested in
evaluating the quality of the obtained abstractions and assumptions,
and the efficiency of our interleaved approach.

\section*{Acknowledgment} 
Corina P\u{a}s\u{a}reanu uses this occasion to remember the wonderful times she  spent at Kansas State University during her graduate studies. She would like to thank Dave Schmidt and the other professors in the Department of Computing and Information Sciences for having started her interest in abstraction and compositional reasoning, which have been the subject of her thesis and a recurrent research theme in her work since then.

\nocite{*}
\bibliographystyle{eptcs}
\bibliography{generic}

\begin{thebibliography}{10}
\providecommand{\bibitemdeclare}[2]{}
\providecommand{\surnamestart}{}
\providecommand{\surnameend}{}
\providecommand{\urlprefix}{Available at }
\providecommand{\url}[1]{\texttt{#1}}
\providecommand{\href}[2]{\texttt{#2}}
\providecommand{\urlalt}[2]{\href{#1}{#2}}
\providecommand{\doi}[1]{doi:\urlalt{http://dx.doi.org/#1}{#1}}
\providecommand{\bibinfo}[2]{#2}

\bibitemdeclare{inproceedings}{DBLP:conf/popl/AlurCMN05}
\bibitem{DBLP:conf/popl/AlurCMN05}
\bibinfo{author}{Rajeev \surnamestart Alur\surnameend}, \bibinfo{author}{Pavol
  \surnamestart Cern{\'y}\surnameend}, \bibinfo{author}{P.~\surnamestart
  Madhusudan\surnameend} \& \bibinfo{author}{Wonhong \surnamestart
  Nam\surnameend} (\bibinfo{year}{2005}): \emph{\bibinfo{title}{Synthesis of
  interface specifications for Java classes}}.
\newblock In \bibinfo{editor}{Palsberg} \& \bibinfo{editor}{Abadi}
  \cite{DBLP:conf/popl/2005}, pp. \bibinfo{pages}{98--109},
  \doi{10.1145/1040305.1040314}.
\newblock \urlprefix\url{http://dl.acm.org/citation.cfm?id=1040305}.

\bibitemdeclare{article}{angluin87}
\bibitem{angluin87}
\bibinfo{author}{Dana \surnamestart Angluin\surnameend} (\bibinfo{year}{1987}):
  \emph{\bibinfo{title}{Learning Regular Sets from Queries and
  Counterexamples}}.
\newblock {\sl \bibinfo{journal}{Inf. Comput.}}
  \bibinfo{volume}{75}(\bibinfo{number}{2}), pp. \bibinfo{pages}{87--106},
  \doi{10.1016/0890-5401(87)90052-6}.

\bibitemdeclare{inproceedings}{DBLP:conf/cav/BeyerHS07}
\bibitem{DBLP:conf/cav/BeyerHS07}
\bibinfo{author}{Dirk \surnamestart Beyer\surnameend},
  \bibinfo{author}{Thomas~A. \surnamestart Henzinger\surnameend} \&
  \bibinfo{author}{Vasu \surnamestart Singh\surnameend} (\bibinfo{year}{2007}):
  \emph{\bibinfo{title}{Algorithms for Interface Synthesis}}.
\newblock In \bibinfo{editor}{Damm} \& \bibinfo{editor}{Hermanns}
  \cite{DBLP:conf/cav/2007}, pp. \bibinfo{pages}{4--19},
  \doi{10.1007/978-3-540-73368-3\_4}.

\bibitemdeclare{inproceedings}{magic}
\bibitem{magic}
\bibinfo{author}{Sagar \surnamestart Chaki\surnameend},
  \bibinfo{author}{Edmund~M. \surnamestart Clarke\surnameend},
  \bibinfo{author}{Alex \surnamestart Groce\surnameend},
  \bibinfo{author}{Somesh \surnamestart Jha\surnameend} \&
  \bibinfo{author}{Helmut \surnamestart Veith\surnameend}
  (\bibinfo{year}{2003}): \emph{\bibinfo{title}{Modular Verification of
  Software Components in C}}.
\newblock In: {\sl \bibinfo{booktitle}{ICSE}}, pp. \bibinfo{pages}{385--395},
  \doi{10.1109/ICSE.2003.1201217}.

\bibitemdeclare{article}{magic2}
\bibitem{magic2}
\bibinfo{author}{Sagar \surnamestart Chaki\surnameend},
  \bibinfo{author}{Jo{\"e}l \surnamestart Ouaknine\surnameend},
  \bibinfo{author}{Karen \surnamestart Yorav\surnameend} \&
  \bibinfo{author}{Edmund~M. \surnamestart Clarke\surnameend}
  (\bibinfo{year}{2003}): \emph{\bibinfo{title}{Automated Compositional
  Abstraction Refinement for Concurrent C Programs: A Two-Level Approach}}.
\newblock {\sl \bibinfo{journal}{Electr. Notes Theor. Comput. Sci.}}
  \bibinfo{volume}{89}(\bibinfo{number}{3}), pp. \bibinfo{pages}{417--432},
  \doi{10.1016/S1571-0661(05)80004-0}.

\bibitemdeclare{proceedings}{DBLP:conf/fase/2009}
\bibitem{DBLP:conf/fase/2009}
\bibinfo{editor}{Marsha \surnamestart Chechik\surnameend} \&
  \bibinfo{editor}{Martin \surnamestart Wirsing\surnameend}, editors
  (\bibinfo{year}{2009}): \emph{\bibinfo{title}{Fundamental Approaches to
  Software Engineering, 12th International Conference, FASE 2009, Held as Part
  of the Joint European Conferences on Theory and Practice of Software, ETAPS
  2009, York, UK, March 22-29, 2009. Proceedings}}. {\sl
  \bibinfo{series}{Lecture Notes in Computer Science}} \bibinfo{volume}{5503},
  \bibinfo{publisher}{Springer}, \doi{10.1007/978-3-642-00593-0}.

\bibitemdeclare{inproceedings}{DBLP:conf/cav/ClarkeGJLV00}
\bibitem{DBLP:conf/cav/ClarkeGJLV00}
\bibinfo{author}{Edmund~M. \surnamestart Clarke\surnameend},
  \bibinfo{author}{Orna \surnamestart Grumberg\surnameend},
  \bibinfo{author}{Somesh \surnamestart Jha\surnameend}, \bibinfo{author}{Yuan
  \surnamestart Lu\surnameend} \& \bibinfo{author}{Helmut \surnamestart
  Veith\surnameend} (\bibinfo{year}{2000}):
  \emph{\bibinfo{title}{Counterexample-Guided Abstraction Refinement}}.
\newblock In \bibinfo{editor}{Emerson} \& \bibinfo{editor}{Sistla}
  \cite{DBLP:conf/cav/2000}, pp. \bibinfo{pages}{154--169},
  \doi{10.1007/10722167\_15}.

\bibitemdeclare{inproceedings}{cobleigh03}
\bibitem{cobleigh03}
\bibinfo{author}{Jamieson~M. \surnamestart Cobleigh\surnameend},
  \bibinfo{author}{Dimitra \surnamestart Giannakopoulou\surnameend} \&
  \bibinfo{author}{Corina~S. \surnamestart P\u{a}s\u{a}reanu\surnameend}
  (\bibinfo{year}{2003}): \emph{\bibinfo{title}{Learning Assumptions for
  Compositional Verification}}.
\newblock In \bibinfo{editor}{Garavel} \& \bibinfo{editor}{Hatcliff}
  \cite{DBLP:conf/tacas/2003}, pp. \bibinfo{pages}{331--346},
  \doi{10.1007/3-540-36577-X\_24}.

\bibitemdeclare{inproceedings}{bandera}
\bibitem{bandera}
\bibinfo{author}{James~C. \surnamestart Corbett\surnameend},
  \bibinfo{author}{Matthew~B. \surnamestart Dwyer\surnameend},
  \bibinfo{author}{John \surnamestart Hatcliff\surnameend},
  \bibinfo{author}{Shawn \surnamestart Laubach\surnameend},
  \bibinfo{author}{Corina~S. \surnamestart Pasareanu\surnameend},
  \bibinfo{author}{\surnamestart Robby\surnameend} \& \bibinfo{author}{Hongjun
  \surnamestart Zheng\surnameend} (\bibinfo{year}{2000}):
  \emph{\bibinfo{title}{Bandera: extracting finite-state models from Java
  source code}}.
\newblock In: {\sl \bibinfo{booktitle}{ICSE}}, pp. \bibinfo{pages}{439--448},
  \doi{10.1145/337180.337234}.

\bibitemdeclare{inproceedings}{DBLP:conf/popl/CousotC77}
\bibitem{DBLP:conf/popl/CousotC77}
\bibinfo{author}{Patrick \surnamestart Cousot\surnameend} \&
  \bibinfo{author}{Radhia \surnamestart Cousot\surnameend}
  (\bibinfo{year}{1977}): \emph{\bibinfo{title}{Abstract Interpretation: A
  Unified Lattice Model for Static Analysis of Programs by Construction or
  Approximation of Fixpoints}}.
\newblock In \bibinfo{editor}{Graham} et~al.  \cite{DBLP:conf/popl/77}, pp.
  \bibinfo{pages}{238--252}, \doi{10.1145/512950.512973}.
\newblock \urlprefix\url{http://dl.acm.org/citation.cfm?id=512950}.

\bibitemdeclare{proceedings}{DBLP:conf/cav/2007}
\bibitem{DBLP:conf/cav/2007}
\bibinfo{editor}{Werner \surnamestart Damm\surnameend} \&
  \bibinfo{editor}{Holger \surnamestart Hermanns\surnameend}, editors
  (\bibinfo{year}{2007}): \emph{\bibinfo{title}{Computer Aided Verification,
  19th International Conference, CAV 2007, Berlin, Germany, July 3-7, 2007,
  Proceedings}}. {\sl \bibinfo{series}{Lecture Notes in Computer Science}}
  \bibinfo{volume}{4590}, \bibinfo{publisher}{Springer}.

\bibitemdeclare{proceedings}{DBLP:conf/cav/2000}
\bibitem{DBLP:conf/cav/2000}
\bibinfo{editor}{E.~Allen \surnamestart Emerson\surnameend} \&
  \bibinfo{editor}{A.~Prasad \surnamestart Sistla\surnameend}, editors
  (\bibinfo{year}{2000}): \emph{\bibinfo{title}{Computer Aided Verification,
  12th International Conference, CAV 2000, Chicago, IL, USA, July 15-19, 2000,
  Proceedings}}. {\sl \bibinfo{series}{Lecture Notes in Computer Science}}
  \bibinfo{volume}{1855}, \bibinfo{publisher}{Springer}.

\bibitemdeclare{proceedings}{DBLP:conf/tacas/2003}
\bibitem{DBLP:conf/tacas/2003}
\bibinfo{editor}{Hubert \surnamestart Garavel\surnameend} \&
  \bibinfo{editor}{John \surnamestart Hatcliff\surnameend}, editors
  (\bibinfo{year}{2003}): \emph{\bibinfo{title}{Tools and Algorithms for the
  Construction and Analysis of Systems, 9th International Conference, TACAS
  2003, Held as Part of the Joint European Conferences on Theory and Practice
  of Software, ETAPS 2003, Warsaw, Poland, April 7-11, 2003, Proceedings}}.
  {\sl \bibinfo{series}{Lecture Notes in Computer Science}}
  \bibinfo{volume}{2619}, \bibinfo{publisher}{Springer}.

\bibitemdeclare{article}{DBLP:journals/fmsd/GiannakopoulouP08}
\bibitem{DBLP:journals/fmsd/GiannakopoulouP08}
\bibinfo{author}{Dimitra \surnamestart Giannakopoulou\surnameend} \&
  \bibinfo{author}{Corina~S. \surnamestart P\u{a}s\u{a}reanu\surnameend}
  (\bibinfo{year}{2008}): \emph{\bibinfo{title}{Special issue on learning
  techniques for compositional reasoning}}.
\newblock {\sl \bibinfo{journal}{Formal Methods in System Design}}
  \bibinfo{volume}{32}(\bibinfo{number}{3}), pp. \bibinfo{pages}{173--174},
  \doi{10.1007/s10703-008-0054-9}.

\bibitemdeclare{inproceedings}{DBLP:conf/fase/GiannakopoulouP09}
\bibitem{DBLP:conf/fase/GiannakopoulouP09}
\bibinfo{author}{Dimitra \surnamestart Giannakopoulou\surnameend} \&
  \bibinfo{author}{Corina~S. \surnamestart P\u{a}s\u{a}reanu\surnameend}
  (\bibinfo{year}{2009}): \emph{\bibinfo{title}{Interface Generation and
  Compositional Verification in JavaPathfinder}}.
\newblock In \bibinfo{editor}{Chechik} \& \bibinfo{editor}{Wirsing}
  \cite{DBLP:conf/fase/2009}, pp. \bibinfo{pages}{94--108},
  \doi{10.1007/978-3-642-00593-0\_7}.

\bibitemdeclare{inproceedings}{DBLP:conf/sas/GiannakopoulouRR12}
\bibitem{DBLP:conf/sas/GiannakopoulouRR12}
\bibinfo{author}{Dimitra \surnamestart Giannakopoulou\surnameend},
  \bibinfo{author}{Zvonimir \surnamestart Rakamaric\surnameend} \&
  \bibinfo{author}{Vishwanath \surnamestart Raman\surnameend}
  (\bibinfo{year}{2012}): \emph{\bibinfo{title}{Symbolic Learning of Component
  Interfaces}}.
\newblock In \bibinfo{editor}{Min{\'e}} \& \bibinfo{editor}{Schmidt}
  \cite{DBLP:conf/sas/2012}, pp. \bibinfo{pages}{248--264},
  \doi{10.1007/978-3-642-33125-1\_18}.

\bibitemdeclare{proceedings}{DBLP:conf/popl/77}
\bibitem{DBLP:conf/popl/77}
\bibinfo{editor}{Robert~M. \surnamestart Graham\surnameend},
  \bibinfo{editor}{Michael~A. \surnamestart Harrison\surnameend} \&
  \bibinfo{editor}{Ravi \surnamestart Sethi\surnameend}, editors
  (\bibinfo{year}{1977}): \emph{\bibinfo{title}{Conference Record of the Fourth
  ACM Symposium on Principles of Programming Languages, Los Angeles,
  California, USA, January 1977}}. \bibinfo{publisher}{ACM}.
\newblock \urlprefix\url{http://dl.acm.org/citation.cfm?id=512950}.

\bibitemdeclare{proceedings}{DBLP:conf/padl/2007}
\bibitem{DBLP:conf/padl/2007}
\bibinfo{editor}{Michael \surnamestart Hanus\surnameend}, editor
  (\bibinfo{year}{2007}): \emph{\bibinfo{title}{Practical Aspects of
  Declarative Languages, 9th International Symposium, PADL 2007, Nice, France,
  January 14-15, 2007}}. {\sl \bibinfo{series}{Lecture Notes in Computer
  Science}} \bibinfo{volume}{4354}, \bibinfo{publisher}{Springer}.

\bibitemdeclare{inproceedings}{blast}
\bibitem{blast}
\bibinfo{author}{Thomas~A. \surnamestart Henzinger\surnameend},
  \bibinfo{author}{Ranjit \surnamestart Jhala\surnameend} \&
  \bibinfo{author}{Rupak \surnamestart Majumdar\surnameend}
  (\bibinfo{year}{2004}): \emph{\bibinfo{title}{Race checking by context
  inference}}.
\newblock In: {\sl \bibinfo{booktitle}{PLDI}}, pp. \bibinfo{pages}{1--13},
  \doi{10.1145/996841.996844}.

\bibitemdeclare{inproceedings}{DBLP:conf/sigsoft/HenzingerJM05}
\bibitem{DBLP:conf/sigsoft/HenzingerJM05}
\bibinfo{author}{Thomas~A. \surnamestart Henzinger\surnameend},
  \bibinfo{author}{Ranjit \surnamestart Jhala\surnameend} \&
  \bibinfo{author}{Rupak \surnamestart Majumdar\surnameend}
  (\bibinfo{year}{2005}): \emph{\bibinfo{title}{Permissive interfaces}}.
\newblock In \bibinfo{editor}{Wermelinger} \& \bibinfo{editor}{Gall}
  \cite{DBLP:conf/sigsoft/2005}, pp. \bibinfo{pages}{31--40},
  \doi{10.1145/1081706.1081713}.

\bibitemdeclare{article}{jones83a}
\bibitem{jones83a}
\bibinfo{author}{Cliff~B. \surnamestart Jones\surnameend}
  (\bibinfo{year}{1983}): \emph{\bibinfo{title}{Tentative Steps Toward a
  Development Method for Interfering Programs}}.
\newblock {\sl \bibinfo{journal}{ACM Trans. Program. Lang. Syst.}}
  \bibinfo{volume}{5}(\bibinfo{number}{4}), pp. \bibinfo{pages}{596--619},
  \doi{10.1145/69575.69577}.

\bibitemdeclare{proceedings}{DBLP:conf/sas/2012}
\bibitem{DBLP:conf/sas/2012}
\bibinfo{editor}{Antoine \surnamestart Min{\'e}\surnameend} \&
  \bibinfo{editor}{David \surnamestart Schmidt\surnameend}, editors
  (\bibinfo{year}{2012}): \emph{\bibinfo{title}{Static Analysis - 19th
  International Symposium, SAS 2012, Deauville, France, September 11-13, 2012.
  Proceedings}}. {\sl \bibinfo{series}{Lecture Notes in Computer Science}}
  \bibinfo{volume}{7460}, \bibinfo{publisher}{Springer},
  \doi{10.1007/978-3-642-33125-1}.

\bibitemdeclare{inproceedings}{DBLP:conf/cav/NamjoshiK00}
\bibitem{DBLP:conf/cav/NamjoshiK00}
\bibinfo{author}{Kedar~S. \surnamestart Namjoshi\surnameend} \&
  \bibinfo{author}{Robert~P. \surnamestart Kurshan\surnameend}
  (\bibinfo{year}{2000}): \emph{\bibinfo{title}{Syntactic Program
  Transformations for Automatic Abstraction}}.
\newblock In \bibinfo{editor}{Emerson} \& \bibinfo{editor}{Sistla}
  \cite{DBLP:conf/cav/2000}, pp. \bibinfo{pages}{435--449},
  \doi{10.1007/10722167\_33}.

\bibitemdeclare{proceedings}{DBLP:conf/popl/2005}
\bibitem{DBLP:conf/popl/2005}
\bibinfo{editor}{Jens \surnamestart Palsberg\surnameend} \&
  \bibinfo{editor}{Mart\'{\i}n \surnamestart Abadi\surnameend}, editors
  (\bibinfo{year}{2005}): \emph{\bibinfo{title}{Proceedings of the 32nd ACM
  SIGPLAN-SIGACT Symposium on Principles of Programming Languages, POPL 2005,
  Long Beach, California, USA, January 12-14, 2005}}. \bibinfo{publisher}{ACM}.
\newblock \urlprefix\url{http://dl.acm.org/citation.cfm?id=1040305}.

\bibitemdeclare{inbook}{pnueli84}
\bibitem{pnueli84}
\bibinfo{author}{A.~\surnamestart Pnueli\surnameend} (\bibinfo{year}{1985}):
  \emph{\bibinfo{title}{In transition from global to modular temporal reasoning
  about programs}}, pp. \bibinfo{pages}{123--144}.
\newblock \bibinfo{publisher}{Springer-Verlag New York, Inc.},
  \bibinfo{address}{New York, NY, USA}, \doi{10.1007/978-3-642-82453-1\_5}.
\newblock \urlprefix\url{http://portal.acm.org/citation.cfm?id=101969.101977}.

\bibitemdeclare{inproceedings}{DBLP:conf/padl/PodelskiR07}
\bibitem{DBLP:conf/padl/PodelskiR07}
\bibinfo{author}{Andreas \surnamestart Podelski\surnameend} \&
  \bibinfo{author}{Andrey \surnamestart Rybalchenko\surnameend}
  (\bibinfo{year}{2007}): \emph{\bibinfo{title}{ARMC: The Logical Choice for
  Software Model Checking with Abstraction Refinement}}.
\newblock In \bibinfo{editor}{Hanus}  \cite{DBLP:conf/padl/2007}, pp.
  \bibinfo{pages}{245--259}, \doi{10.1007/978-3-540-69611-7\_16}.

\bibitemdeclare{article}{thesis}
\bibitem{thesis}
\bibinfo{author}{Corina~S. \surnamestart P\u{a}s\u{a}reanu\surnameend}
  (\bibinfo{year}{2001}): \emph{\bibinfo{title}{Abstraction and Modular
  Reasoning for the Verification of Software}}.
\newblock {\sl \bibinfo{journal}{PhD Thesis, Kansas State University}}.

\bibitemdeclare{article}{DBLP:journals/fmsd/PasareanuGBCB08}
\bibitem{DBLP:journals/fmsd/PasareanuGBCB08}
\bibinfo{author}{Corina~S. \surnamestart P\u{a}s\u{a}reanu\surnameend},
  \bibinfo{author}{Dimitra \surnamestart Giannakopoulou\surnameend},
  \bibinfo{author}{Mihaela~Gheorghiu \surnamestart Bobaru\surnameend},
  \bibinfo{author}{Jamieson~M. \surnamestart Cobleigh\surnameend} \&
  \bibinfo{author}{Howard \surnamestart Barringer\surnameend}
  (\bibinfo{year}{2008}): \emph{\bibinfo{title}{Learning to divide and conquer:
  applying the {L}* algorithm to automate assume-guarantee reasoning}}.
\newblock {\sl \bibinfo{journal}{Formal Methods in System Design}}
  \bibinfo{volume}{32}(\bibinfo{number}{3}), pp. \bibinfo{pages}{175--205},
  \doi{10.1007/s10703-008-0049-6}.

\bibitemdeclare{article}{rivest93}
\bibitem{rivest93}
\bibinfo{author}{Ronald~L. \surnamestart Rivest\surnameend} \&
  \bibinfo{author}{Robert~E. \surnamestart Schapire\surnameend}
  (\bibinfo{year}{1993}): \emph{\bibinfo{title}{Inference of Finite Automata
  Using Homing Sequences}}.
\newblock {\sl \bibinfo{journal}{Inf. Comput.}}
  \bibinfo{volume}{103}(\bibinfo{number}{2}), pp. \bibinfo{pages}{299--347},
  \doi{10.1006/inco.1993.1021}.

\bibitemdeclare{inproceedings}{cav2010}
\bibitem{cav2010}
\bibinfo{author}{Rishabh \surnamestart Singh\surnameend},
  \bibinfo{author}{Dimitra \surnamestart Giannakopoulou\surnameend} \&
  \bibinfo{author}{Corina~S. \surnamestart P\u{a}s\u{a}reanu\surnameend}
  (\bibinfo{year}{2010}): \emph{\bibinfo{title}{Learning Component Interfaces
  with May and Must Abstractions}}.
\newblock In \bibinfo{editor}{Touili} et~al.  \cite{DBLP:conf/cav/2010}, pp.
  \bibinfo{pages}{527--542}, \doi{10.1007/978-3-642-14295-6\_45}.

\bibitemdeclare{proceedings}{DBLP:conf/cav/2010}
\bibitem{DBLP:conf/cav/2010}
\bibinfo{editor}{Tayssir \surnamestart Touili\surnameend},
  \bibinfo{editor}{Byron \surnamestart Cook\surnameend} \&
  \bibinfo{editor}{Paul \surnamestart Jackson\surnameend}, editors
  (\bibinfo{year}{2010}): \emph{\bibinfo{title}{Computer Aided Verification,
  22nd International Conference, CAV 2010, Edinburgh, UK, July 15-19, 2010.
  Proceedings}}. {\sl \bibinfo{series}{Lecture Notes in Computer Science}}
  \bibinfo{volume}{6174}, \bibinfo{publisher}{Springer},
  \doi{10.1007/978-3-642-14295-6}.

\bibitemdeclare{proceedings}{DBLP:conf/sigsoft/2005}
\bibitem{DBLP:conf/sigsoft/2005}
\bibinfo{editor}{Michel \surnamestart Wermelinger\surnameend} \&
  \bibinfo{editor}{Harald \surnamestart Gall\surnameend}, editors
  (\bibinfo{year}{2005}): \emph{\bibinfo{title}{Proceedings of the 10th
  European Software Engineering Conference held jointly with 13th ACM SIGSOFT
  International Symposium on Foundations of Software Engineering, 2005, Lisbon,
  Portugal, September 5-9, 2005}}. \bibinfo{publisher}{ACM}.

\end{thebibliography}
\end{document}